# Fold-switching proteins push the boundaries of conformational ensemble prediction

Myeongsang Lee[1] and Lauren L. Porter[1,2,*]

[1]Computational Biology Branch, Division of Intramural Research, National Library of Medicine, National Institutes of Health, Bethesda, MD 20894

[2]Biochemistry and Biophysics Center, National Heart, Lung, and Blood Institute, National Library of Medicine, National Institutes of Health, Bethesda, MD 20892

Email address: myeongsang.lee@nih.gov, porterll@nih.gov

Running title: Fold-switching Proteins and Predicting Protein Conformational Distributions

*Correspondence:

National Library of Medicine

8600 Rockville Pike

Bethesda, MD 20894

porterll@nih.gov

**Abstract**

A protein's function depends critically on its conformational ensemble, a collection of energy weighted structures whose balance depends on temperature and environment. Though recent deep learning (DL) methods have substantially advanced predictions of single protein structures, computationally modeling conformational ensembles remains a challenge. Here, we focus on modeling fold-switching proteins, which remodel their secondary and/or tertiary structures and change their functions in response to cellular stimuli. These underrepresented members of the protein universe serve as test cases for a method's generalizability. They reveal that DL models often predict conformational ensembles by association with training-set structures, limiting generalizability. These observations suggest use cases for when DL methods will likely succeed or fail. Developing computational methods that successfully identify new fold-switching proteins from large pools of candidates may advance modeling conformational ensembles more broadly.

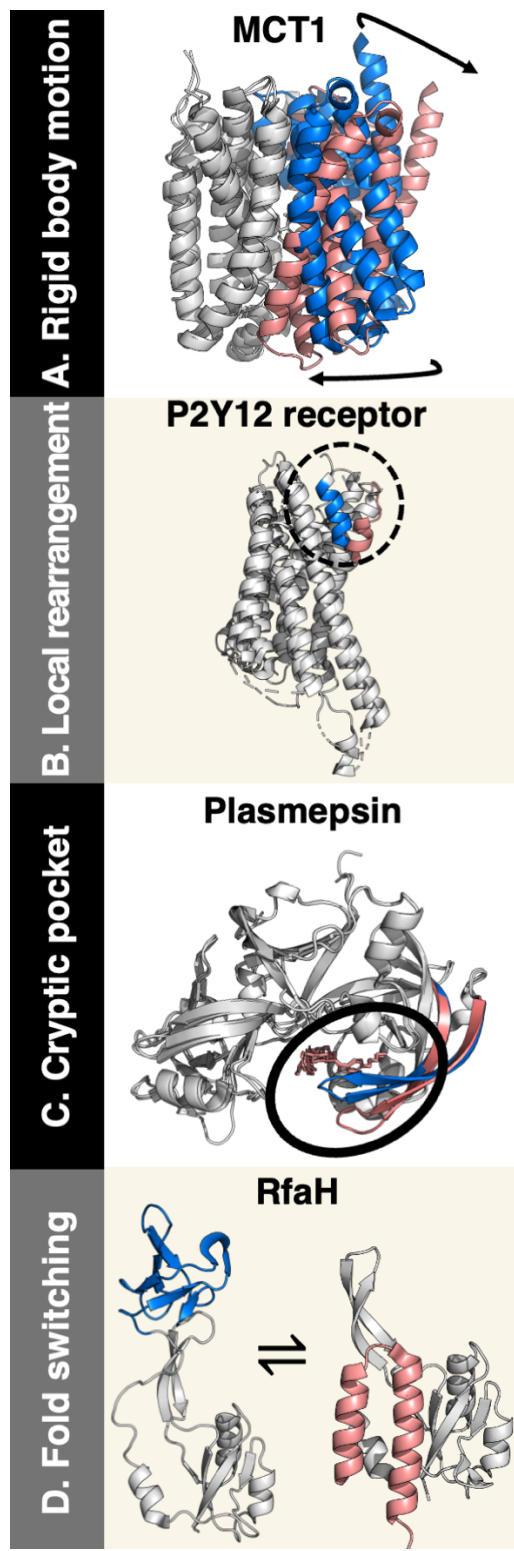

## 1. Introduction

Since the concept of the protein universe was first introduced (1), over 350 million protein sequences have been deposited in the UniProt database of curated protein sequences (2), and over 3.1 billion metagenomic protein sequences have been made publicly available (3). Within this vast characterized sequence space, each protein populates an ensemble of energy-weighted conformations. Experimental methods, such as X-ray crystallography, tend to characterize the most energetically favorable (and therefore most populated) conformation of a folded protein. Still, less populated alternative conformations play important roles in protein function (4) and dysfunction related to human disease (5, 6). Thus, modeling protein ensembles has been an important aim in computational biology, successfully revealing mechanistic details of protein function in some cases (7-9).

The diverse conformations within a globular protein's structural ensemble typically involve some combination of domain reorientation and local conformational rearrangement (**Figure 1**). For instance, dimeric membrane transport proteins often undergo large interdomain reorientations when they transition between outward- and inward-facing conformations enabling cellular influx and efflux, as exemplified by the human monocarboxylate transporter 1 (MCT1, **Figure 1A**) (10). Cellular signaling proteins known as G-protein-coupled receptors (GPCRs) respond to numerous extracellular signals through both rigid body reorientations and local conformational rearrangements, as exemplified by $P2Y_{12}$, an ADP-binding GPCR commonly subtle local conformational rearrangements, as exemplified by the malaria protein plasmepsin (**Figure 1C**). These rearrangements can affect protein function through long-range interactions (allostery) (15). Cryptic pockets have been recognized as important targets for drug discovery (16). Though not globular, intrinsically disordered proteins (IDPs) can adopt numerous diverse conformations important for different cellular functions. Excellent reviews describing IDPs in detail can be found elsewhere (17, 18).

**Figure 1. Proteins undergo different sorts of conformational changes.** A. The inward-facing (blue) and outward-facing (pink) conformations of human monocarboxylate transporter differ by a rigid body reorientation of their C-terminal lobe (marked with arrows). B. The apo inactive form of the human GPCR P2Y12 (blue) undergoes local helix unwinding (pink) upon activation by GDP. C. The malaria protein plasmepsin II has a cryptic pocket able to bind some small molecules (pink), in contrast to its unbound form which occludes this binding pocket (blue). D. The C-terminal domain of bacterial RfaH switches from its activated β-roll fold (blue) into an autoinhibited helical form (pink) in the absence of RNA polymerase and *operon polarity suppressor* DNA. In all four cases, conformational changes are highlighted in blue for dominant AlphaFold2 predictions and pink for alternative; the rest of the protein is colored gray. All protein figures were made with PyMOL (123).

Unlike the ensembles of typical globular proteins, whose secondary structures remain largely fixed when conformational changes occur, fold-switching proteins interconvert between diverse conformations with distinct secondary structures, allowing them to perform different functions in response to their environments (19, 20). For instance, the C-terminal domain (CTD) of the transcription regulator RfaH undergoes a reversible α-helix <-> β-sheet transition in response to binding both RNA polymerase and the *operon polarity suppressor* (*ops*) DNA sequence (**Figure 1D**) (21). This dramatic secondary structure remodeling confines RfaH's transcription and translation regulation to *ops* DNA and enables efficient translation (22); a recent computational and experimental analysis suggests that fold switching is conserved among many diverse RfaH homologs—up to 25% of the universally conserved NusG transcription regulator family of which it is a part (23).

Here, we focus on recent approaches used to model fold-switched conformations of proteins and what their outcomes reveal about the state-of-the-art. Though fold switchers constitute a relatively small fraction of the protein universe, perhaps 4-5% (20, 24, 25), their rarity offers an advantage when evaluating predictive models. Artificial intelligence (AI)-based protein structure predictors have struggled to learn generalizable rules for the data-poor fold switching proteome. Instead, they readily predict a single conformation of fold switchers while often missing experimentally characterized alternatives (26-28). This lack of generalizability provided the first evidence that AI-based models predict some protein structures from memorization rather than a robust learning of folding physics (29-32). Indeed, AI-models failed to predict important conformational changes in the most recent Critical Assessment of techniques for protein Structure Prediction (CASP) (33) and have failed to generalize in other tasks such as binding small molecules and peptides (34-36), and modeling physical features of proteins, such as sidechain-sidechain interactions (37). Physically-based simulations, such as molecular dynamics (MD), have been used to model conformational ensembles of fold switchers with some success (8, 38, 39), though only when their two most populated conformations were known. Thus, predicting new alternative conformations of fold switchers remains an outstanding challenge. Since all proteins are subject to the same laws of physics, addressing this challenge could inform more robust methods for protein structure prediction and ensemble generation.

## 2. Predicting fold switching with AI-based models

## 2.1 Deep learning models readily predict dominant conformations but often struggle with alternative conformations

AlphaFold–a Nobel prizewinning deep learning model–revolutionized protein structure prediction by generating models of protein structures that are often highly accurate (40, 41). These models are generated from the amino acid sequence of a target protein and a **multiple sequence alignment (MSA)**, a collection of homologous sequences with conserved amino acids in register. The structural models produced by AlphaFold2 can often predict protein-protein interactions with high accuracy as well. Furthermore, recently developed models such as AlphaFold3, Chai-1 (42), RoseTTAfold All-Atom (43), and Boltz-2 (44), predict structures of proteins bound to small molecules, DNA, and RNA as well. These **co-folding models** generate the structures of proteins and their binders together rather than generating them separately and then docking the binder to the protein of interest. Together, these deep learning models have greatly expanded the modeler's toolkit for protein structure prediction and design.

Despite the advances that deep learning models have enabled in predicting single protein structures, they often struggle to predict **protein ensembles**: energy weighted distributions of protein conformations (45). Instead, they tend to predict single conformations of proteins (26), which we term **dominant conformations**. Most dominant conformations correspond the conformation most represented in the training set of a deep learning model (32), though some correspond to less common structures memorized during training instead (29, 46). By contrast, deep learning models often struggle to predict **alternative conformations**, which correspond to any conformation other than dominant. This section will focus primarily on predictions of experimentally characterized alternative conformations. Though these alternative conformations do not represent the full protein ensemble, their experimental characterization suggests that they are likely lower energy members of the ensemble, and the ensemble would not be predicted fully without them.

**Pair representations**–probability distributions of pairwise distances between amino acids used by deep learning models to infer protein structures–appear to limit the sorts of alternative conformations that are predicted readily (47). These representations are inferred from MSAs inputted into deep-learning models or evolutionary couplings learned from protein language models (PLMs), such as ESM-3 (48). PLMs are trained for broad tasks, such as recognizing evolutionary information and inferring function (49) These broad training objectives–though useful in many applications–can limit performance of specific structural tasks such as generating diverse conformers based from the same sequence input. For example, ESM-2 disables **dropout**–ignoring units of the neural network, which can increase the variability of predicted outputs–during inference, yielding more homogeneous structure predictions. Further, sequence **embeddings** (multidimensional vector representations of a sequence input) are fixed during ESM-3 inference, leading to deterministic structure predictions (48, 50). Some alternative conformations arising from rigid body motions and/or local conformational rearrangements have pairwise representations very similar to dominant (**Figure 2A**). It is straightforward to imagine how alternative conformations arising from such similar representations might be sampled stochastically from the same input MSA used to generate dominant conformations (48, 51), though this does not always occur (30).

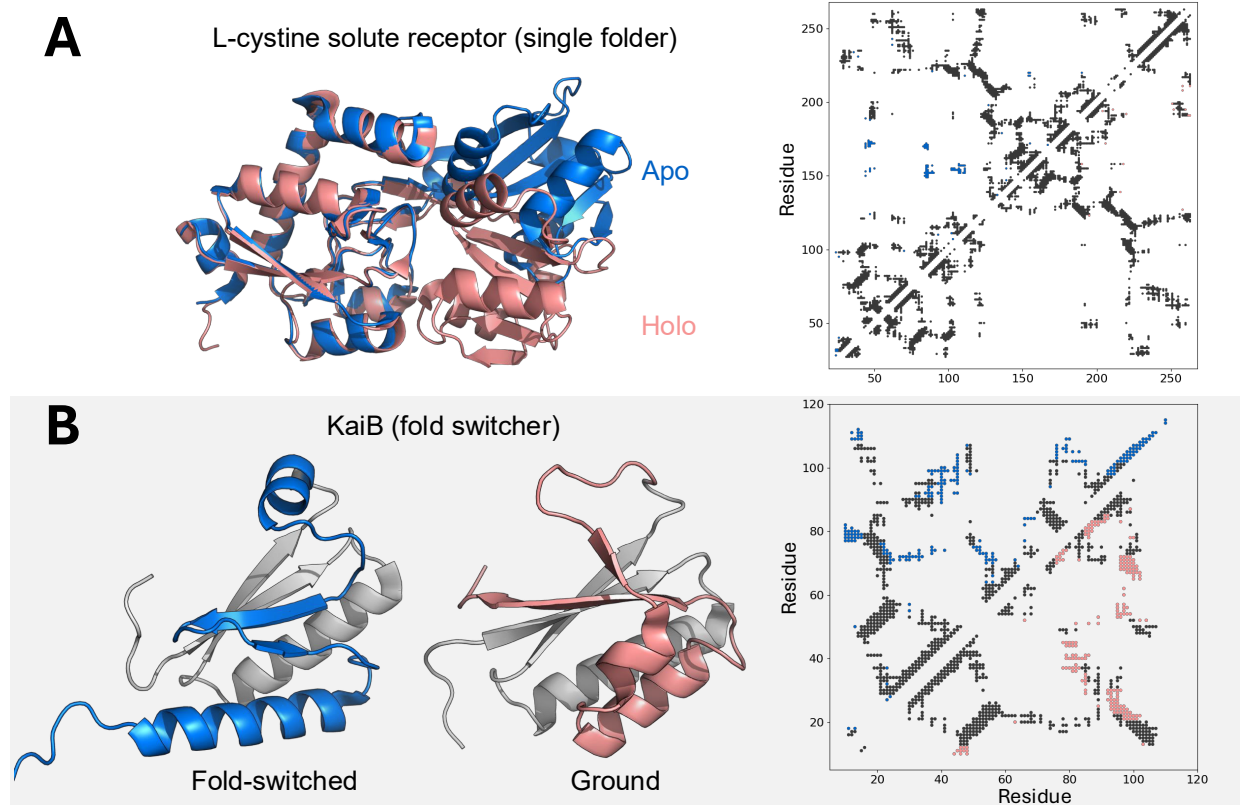

**Figure 2.** While single folders tend to undergo conformational changes that largely preserve their residue-residue contacts, fold switchers can interconvert between conformations with substantially different contacts. **A.** The apo (blue) and holo (pink) forms of L-cystine solute receptor have different conformations, but their overall residue-residue contact maps (left) are largely similar. Upper/lower diagonals of contact map correspond to apo/holo structures; black contacts are common to both conformations; contacts unique to apo/holo are blue/pink. **B.** By contrast, the fold switched and ground conformations of KaiB have very different secondary structures with distinct contact maps. Upper/lower diagonals of contact map correspond to fold-switched/ground structures; black contacts are common to both conformations; contacts unique to fold-switched/ground are blue/pink. The fold-switching regions of KaiB are colored blue and pink while the part of the fold that does not switch is gray. For both panels, blue/pink represent the dominant/alternative conformations predicted by AlphaFold. Ribbon diagrams were made with PyMOL (123). Contact maps were made with a heavy atom distance cutoff of 8 Å.

Fold-switching proteins challenge deep learning models because their dominant and alternative conformations can have very different pair representations (**Figure 2B**). It is less clear how disparate conformations can result from network dropout and/or stochastic sampling of the same input MSA when its overall coevolutionary signal corresponds well with one conformation but not the other. Approaches such as MSA subsampling (24, 52), sequence clustering (53), and random MSA column knockouts with (51) or without network dropout (54) have been developed to overcome this barrier. Nevertheless, none of these approaches consistently and accurately predict alternative conformations of fold-switching proteins, nor do co-folding models (30).

A recent benchmarking of AlphaFold-based methods on fold-switching proteins suggested that successful predictions arise from memorization of training set structures (30). These methods included proposed methods for MSA modification, the AF2 multimer model, and AF3 with all possible binding partners. Upon sampling ~300,000 structures from 92 fold switchers with alternative conformations in AF's training set, a weak success rate of 35% of was achieved. AF's performance on targets outside of its training set was weaker, with a success rate of 1/7. By examining the AF2 network's behavior on fold-switching proteins from various perspectives, it was concluded that AF2 successfully predicts alternative conformations by associating input MSAs with structures or substructures memorized during training. Supporting this conclusion, AF2 predicted several alternative conformations from single sequences (no coevolutionary information from MSAs) and 0 recycles (one pass through the model, forcing "first impression" predictions). Further, AF2 also predicted structures inconsistent with strong coevolutionary signals present in the input MSA, indicating memorization rather than coevolutionary inference. Subsequent studies suggested memorization of other types of conformational changes (32, 55).

To address the question of how well the AlphaFold2 architecture can predict alternative conformations outside of its training set, Bryant and Noé retrained it on a conformational split of the PDB excluding alternative conformations and templates (47). This retrained version of AlphaFold2, CFold, predicted 57% of alternative conformations correctly. While successes corresponded to small-to-moderate conformational differences–including many local conformational rearrangements and domain reorientations (**Figures 1A-C, 2A**)–CFold systematically struggled to predict large conformational changes. Consistently, when run on fold-switching proteins whose alternative conformations were not in its training set, CFold systematically failed (31, 46).

Deep learning models with other architectures similarly struggle to predict alternative conformations of fold switchers. For instance, the diffusion-based model EigenFold performs well on local conformational rearrangements but poorly on fold switchers (27). A masked language modeling approach also leads to poor predictions of fold switchers (28). Very recently, Apple released SimpleFold (56), a generative AI model for predicting protein structures and alternative conformations with a simpler architecture than AlphaFold that does not use MSAs for structural inference. Previous work indicates that benchmarking fold switchers requires focus on the region that switches folds (**Figure 1**) in addition to the overall protein structure (26, 30). SimpleFold benchmarks focused on the overall protein structure only. Consequently, closer inspection revealed that it failed to predict alternative conformations of fold switchers such as KaiB and Mad2, which AlphaFold-based models can predict.

## 2.2 Coevolutionary inference does not appear to drive AlphaFold-based predictions of alternative conformations

AlphaFold's ability to predict dominant folds outside of its training set has been ascribed to coevolutionary inference (57), or its ability to recognize evolutionary couplings (ECs) from input MSAs. The identities of evolutionarily coupled amino acids are correlated over evolutionary history (58), and ~98% of evolutionarily coupled amino acids are close in space (59). Thus, ECs can inform the pairwise representations AlphaFold uses to predict protein structures (60), and diverse structures based on minor variations in the set of ECs can often be generated readily without further training (**Figure 2A**, (47)). Many fold-switched conformations have very different conformations, and, consequently, very different ECs (**Figure 2B**). If AF structure predictions are primarily dictated by coevolutionary inference, appropriately sampled MSAs could, in principle, enable AlphaFold to predict disparate fold-switched conformations without having encountered them during training. Unfortunately, AlphaFold does not yet appear to perform robust coevolutionary inference at this level (31, 47), relying more on highly represented structures encountered during training (32). Nor does it predict novel alternative conformations consistently and accurately, as indicated by a recent CASP competition (33).

Recently, Wayment-Steele and colleagues argued that AF2 predicts alternative conformations by inferring evolutionary couplings from MSAs clustered by sequence similarity (53, 61). They found that AF2 predicted experimentally consistent conformations for both forms of RfaH and Mad2 (**Figure 3A**) at least 50% of the time from 12/345 (4%) of their clusters. When they shuffled the columns of those clusters, AF2 predicted experimentally consistent conformations less frequently (61). Since column shuffling is assumed to reduce or eliminate evolutionary couplings from input MSAs (62), they concluded that the ECs in their unshuffled clusters were important for successful predictions.

Our observations indicate that Wayment-Steele et al.'s analysis misses the point. The important question is whether AlphaFold requires prior knowledge of an alternative conformation to predict it from a sequence cluster. Findings from several groups answer a resounding yes (30-32, 47), and other groups have found AlphaFold requires prior knowledge to predict other protein properties such as some binding interactions (35, 36). We now substantiate the requirement for prior knowledge further. First, if different coevolutionary signals are necessary for predicting different protein structures, an appreciable difference would be observed between the sequence embeddings used to generate different conformations of the same protein. Instead, the CFold study found that diverse conformations resulted from very similar embeddings (47). Second, if AF2 has learned how to infer coevolution in general–rather than specific ECs or conservation patterns learned implicitly from structures during training–we would expect the AF2 architecture to infer any coevolutionary pattern effectively regardless of what's in its training set and leverage it to predict the appropriate corresponding structure. We observe the opposite. As mentioned previously, CFold is a version of the AF2 architecture that has essentially no ability to memorize structures in its training set and therefore relies on evolutionary couplings from input MSAs to predict protein structure (47). Importantly, the alternative conformations of fold switchers such as RfaH and Mad2 are not in its training set. Though CFold predicted the dominant conformations of several fold-switching proteins from deep MSAs, it failed to predict alternative conformations from the sequence clusters highlighted in Wayment-Steele et al.'s original work (31). This highlights the importance of the training set in predictions of alternative conformations.

Similarly, the supposed evolutionary couplings from Wayment-Steele et al.'s 12 sequence clusters (61) were not sufficient to produce the expected conformations of RfaH and Mad2 using CFold (**Figure 3**). Out of 1200 structures sampled (100/cluster), CFold failed to produce any predictions consistent with experiment. Further, CFold produced predictions contradicting this analysis of 12 clusters:

- The cluster that came closest to predicting an experimentally consistent conformation (RfaH_000) was claimed to contain evolutionary couplings for the helical autoinhibited form of RfaH (**Figure 3A**); CFold produced structures with mostly β-sheet and one short helix instead **Figure 3B1**; this conformation has not been observed experimentally (63, 64).
- One of the two clusters that produced a few helical bundle C-terminal domains (CTDs), RfaH_005, contradicts Wayment-Steele et al.'s claim that it harbors coevolutionary signal for the active β-sheet form (**Figure 3B2,3C**). All structures produced by both clusters lacked the properly folded N-terminal domain (NTD) necessary to stabilize the autoinhibitory helical bundle (63) (**Figure 3B2, 3C**). CFold also predicts a helical bundle from the single sequence of RfaH's CTD, indicating that coevolutionary information is not required to predict this conformation.
- None of the RfaH clusters produced conformations consistent with RfaH's β-sheet active form, though four were expected to (**Figure 3D**). In fact, the closest (RfaH_000) was claimed to harbor evolutionary couplings for the helical autoinhibited form of RfaH (**Figure 3B3**).
- AF2 predicted autoinhibited RfaH from RfaH_049 through means other than coevolutionary inference (61); CFold produced structures with accuracies similar to RfaH_049 from all other clusters except RfaH_000 (**Figures 3C,D**), again arguing against coevolution as the primary driver of these predictions.
- None of the three clusters claimed to have evolutionary couplings for either conformation of Mad2 led to an experimentally consistent structure (**Figure 3B4,5; D&E**).

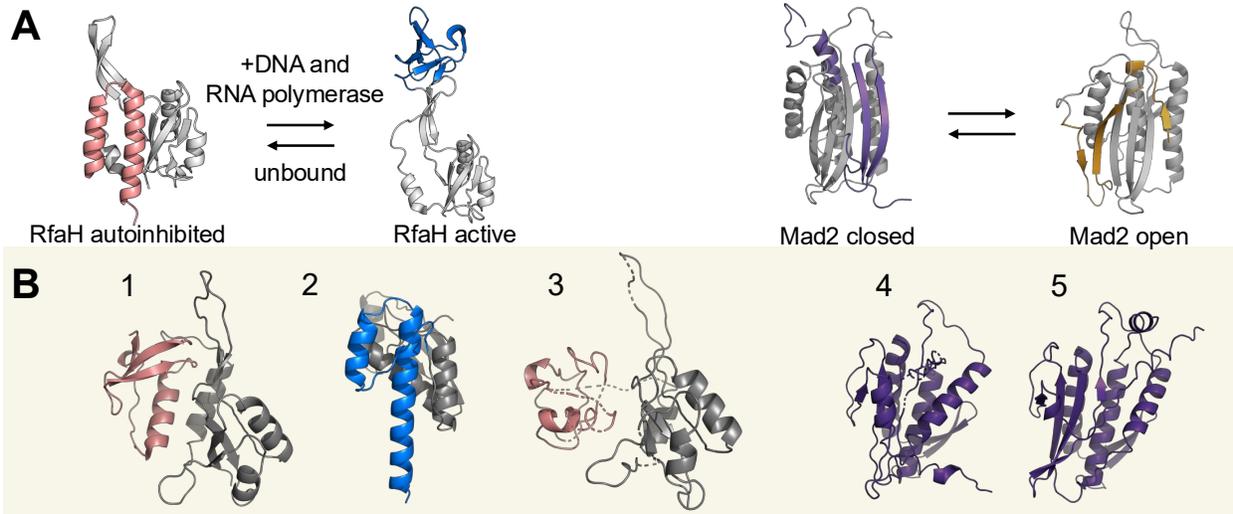
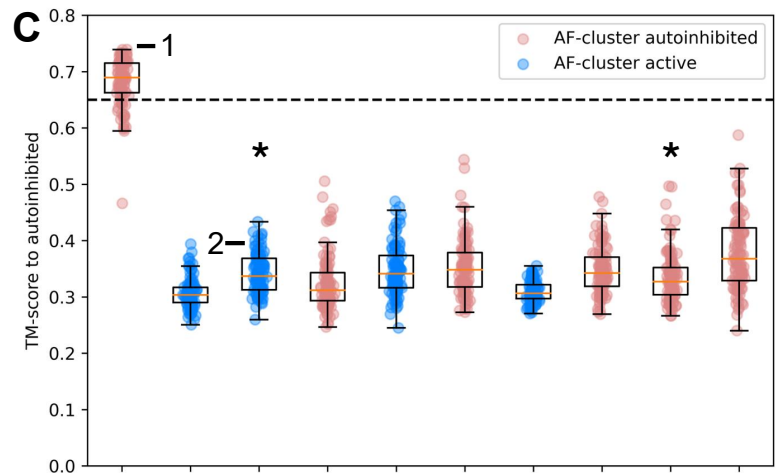
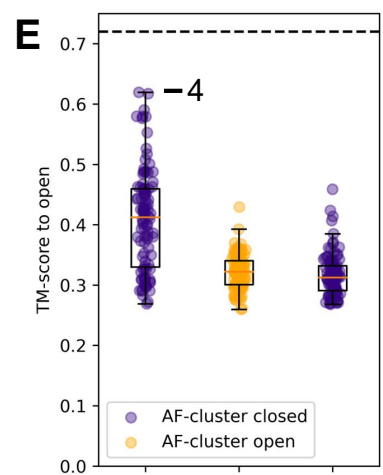
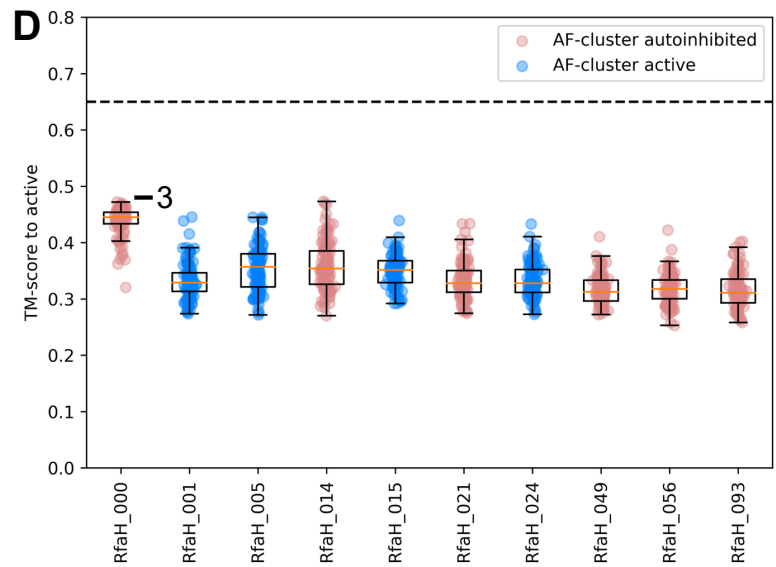
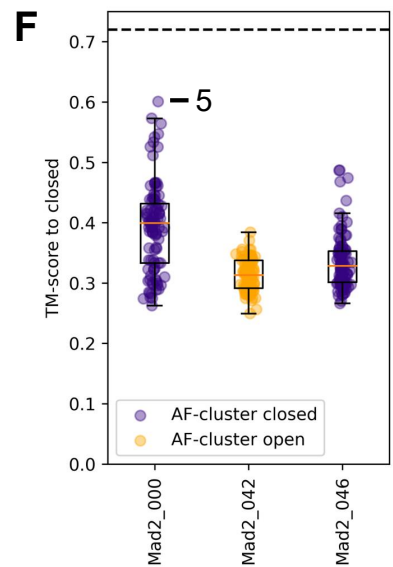

Caption overleaf.

**Figure 3. Evolutionary couplings from sequence clusters do not engender accurate structures of fold switchers**. (A). Experimentally determined structures for the autoinhibited and active forms of RfaH, whose fold switch is triggered by binding both operon polarity suppressor (*ops*) DNA and RNA polymerase. Fold-switching C-terminal domains (CTDs) of the autoinhibited and active forms are colored pink and blue, respectively; the single-folding N-terminal domain (NTD) required to stabilize the autoinhibited conformation is gray. Mad2 interconverts between closed and open forms; the parts of the structure that differ are colored purple and orange, respectively; single-folding regions are gray. (B). CFold predicts structures inconsistent with the claim that AF2 predicts fold-switched conformations from coevolutionary inference of sequence clusters. (1) A cluster claimed to harbor evolutionary couplings consistent with RfaH's helical autoinhibited form produces a mostly β-sheet form with a short stretch of helix instead. (2). A cluster claimed to have evolutionary couplings for the active β-sheet form of RfaH produces 8 helical CTDs with improperly folded NTDs and no CTDs consistent with the active β-sheet form of RfaH. (3). None of the clusters predicted CTDs consistent with the experimentally determined active form of RfaH; the closest was RfaH_000, which was claimed to harbor evolutionary couplings for the autoinhibited form. (4,5). None of the Mad2 sequence clusters engendered experimentally consistent structures. All numbered conformations are annotated on subsequent boxplots. **C-F.** CFold predictions from sequence clusters do not produce experimentally consistent structures. Each boxplot reports 100 structures generated by CFold. Those referenced against autoinhibited RfaH (**C**) use the full-length structure (5OND, chain A) because the autoinhibited conformation requires a properly folded NTD (63); those referenced against active (**D**) use the β-sheet CTD only (6C6S, chain D) because it can form in the absence of NTD. For both **C** and **D**, pink/blue distributions indicate that AF2 predicted ≥50% of structures to be active/autoinhibited from a given sequence cluster. Both the open (**E**) and closed (**F**) forms of Mad2 are referenced against their full-length structures (1S2H and 1DUJ, respectively). For both **E** and **F**, orange/purple distributions indicate that AF2 predicted ≥50% of structures to be open/closed from their respective clusters. Black dotted lines represent accuracy thresholds proposed by Wayment-Steele, et al. when evaluating these 13 clusters (61). Although RfaH_000 is above the threshold for autoinhibited RfaH, it never yielded an experimentally consistent helical hairpin. All protein figures were made with PyMOL (123).

Finally, it should be noted that CFold predicts experimentally consistent structures of active RfaH and closed Mad2 from full MSAs (31), indicating that robust coevolutionary inference is possible. For these reasons, we stand by our original conclusion that coevolutionary inference does not drive AF-cluster based predictions of alternative conformations of fold switchers (29). Rather, AF-cluster's success relies on exposure to training set structures, limiting its predictive scope and explaining its weak performance on a larger fold-switching benchmark (30).

Though a generalizable method for predicting alternative conformations would be preferred above all others, AlphaFold2's ability to associate sequences with structural features learned during training was recently used to good advantage to predict alternative conformations. A method called CF-random leveraged this **sequence association** to maximize the number of alternative conformations it predicted while minimizing the number of structures that needed to be sampled (24, 29). CF-random outperformed other AF-based methods for predicting local conformational rearrangements and domain reorientations. It also predicted fold switchers more robustly than all other AF-based methods than had been benchmarked previously.

### 2.3 Limitations of AlphaFold-based predictions of alternative conformations

Since input MSAs do not generally appear to supply an adequate generative basis for AF-based predictions of disparate alternative conformations, there are limits to what it will typically predict. As discussed previously, AF2 predicts structural fluctuations that fall within the distribution of its dominant coevolutionary signal (**Figure 2A**) (47). This allows for it to sample many rigid body motions and local

# A. AlphaFold sometimes succeeds

## 1. Stochastic sampling
Lysine acetyltransferase

## 2. Sequence association
Sa1 V90T

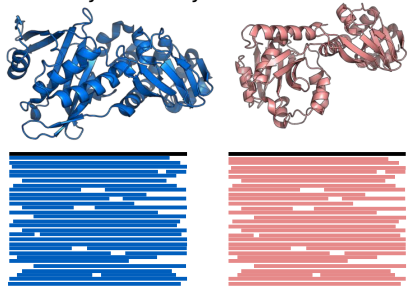
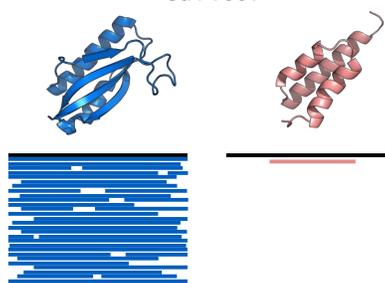

## 3. Contact redistribution

Human interleukin 6

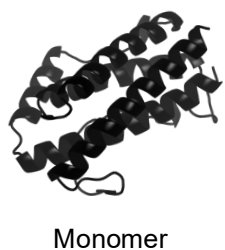

Monomer

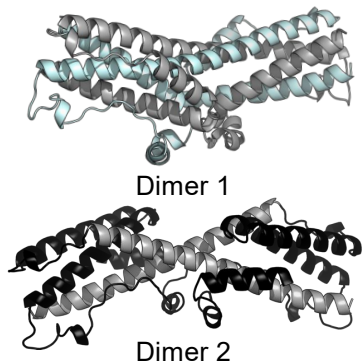

Dimer 1

Dimer 2

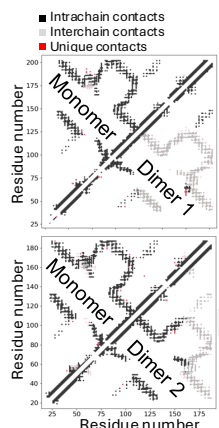

# B. AlphaFold likely fails

## 1. Homologs with new contact patterns
Pro-interleukin 18

## 2. New subunit orientations
MP20

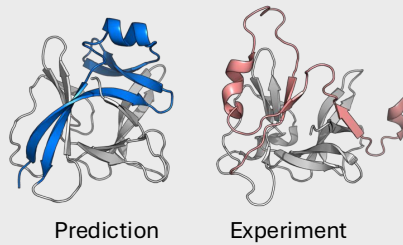

Prediction   Experiment

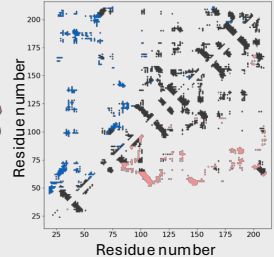

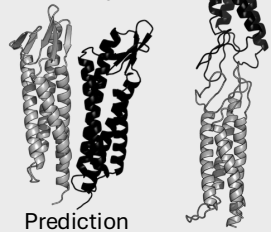

Experiment

Prediction

**Figure 4. Cases in which AlphaFold is more likely to succeed or fail.** **A.** Stochastic MSA sampling with AlphaFold is more likely to succeed when conformational differences are largely consistent with the same contact map, such as for lysine acetyltransferase. MSA subsampling at very shallow depths or pairwise alignments can engender successful predictions of alternative conformations in the training set through sequence association. Furthermore, residue-residue contacts can be exchanged from intrachain to interchain as in the case of monomeric human interleukin-6 and its two domain-swapped forms. Dual-fold contact maps (middle right) both show monomeric intrachain contacts on the upper left diagonal (black). The lower left diagonals show the contact maps of Dimer 1 (top) and Dimer 2 (bottom). While the overall forms of all 3 contact maps are the same, Dimer 1 comprises nearly all interchain contacts through domain swapping (monomeric units colored gray and light cyan to illustrate the extent of interchain contacts), and Dimer 2 comprises a combination of interchain (gray) and intrachain (black) contacts. **B.** AlphaFold will likely fail for protein homologs with structures outside the training set and residue-residue contacts very different from the dominant coevolutionary signal in MSA. This is exemplified by the human protein pro-interleukin-18, whose structure and overall contact patterns differ substantially from its training-set homolog human interleukin 18. AlphaFold mistakenly predicts the structure of human interleukin 18 for pro-interleukin-18 (bottom left image, with structurally distinct region colored blue). The experimentally determined structure of pro-interleukin-18 is shown to its right (structurally distinct region pink). The dual fold contact map of the predicted (upper diagonal) and experimentally determined (lower diagonal) structures have very different contact patterns. AlphaFold also sometimes fails to predict new subunit orientations, as exemplified by the human eye lens protein MP20. It mistakenly predicts a previously observed side-by-side orientation rather than a previously observed stacked orientation; subunits are colored gray and black, respectively. Structures colored blue/pink correspond to dominant/alternative conformations. Ribbon diagrams were made with PyMOL (123). Contact maps were made with a heavy atom distance cutoff of 8 Å.

conformational rearrangements (**Figure 4A**). It can also associate sequences with alternative training set structures. This was observed for the fold-switching protein Sa1 V90T, which AF2 predicts to fold into an α/β plait when using its full MSA as input (65). However, when inputting a pairwise alignment between the sequences of Sa1 V90T and a homolog assuming its alternative 3-α-helix bundle conformation, AF2 predicts that Sa1 V90T assumes the 3-α-helix bundle (24) (**Figure 4B**). Thus, AlphaFold can make new associations between target sequences and training set structures. Associations with protein subdomains can also be made. AF2 and AF3 can also redistribute evolutionary couplings to predict alternative structures. Recently solved structures of human interleukin 16 showed two distinct domain-swapped forms. Though numerous AlphaFold-based methods did not predict either conformation (66), CF-random predicted both successfully (**Figure 4C**). Importantly, the monomeric and both domain-swapped conformations have essentially the same residue-residue contacts (**Figure 4C**) while the distributions of intra- and inter-chain contacts differ between the three structures. It should also be noted, however, that contact redistribution can sometimes produce unphysical structures as was observed for an incorrectly predicted dimeric conformation of XCL1 (30).

Though a substantial number of alternative conformational changes will likely fall into the three categories above, AlphaFold will sometimes fail in other cases. For instance, AlphaFold can mistakenly predict that sequences assume the same structures as their training set homologs when they do not. For instance, both AF2 and AF3 predict that the structure of pro-interleukin-18 is identical to its training-set homolog, interleukin-18 (67). In fact, their structures and contact maps differ substantially (**Figure 4D**), and no structure like pro-interleukin-18 is in AF's training set. Accordingly, we could not generate an accurate structure of pro-interleukin-18 with any AF-based sampling method. A similar result was found for the human cancer isoform BCCIPα (68). Further, MP20 has a stacked oligomer conformation previously unobserved in the PDB (69). Again, we could not get any AF-based sampling method to predict its experimentally observed conformation; a side-by-side conformation is preferentially modeled instead (**Figure 4E**). In short, AlphaFold and other deep learning models have limited ability to predict conformations and interactions outside of their training sets (30, 35). The full scope of these limitations is still unfolding.

**4.4. Mechanistic reasons for predictive limitations of DL models**

Though increasing limitations of AI-based protein ensemble predictions have been observed through trial and error, it is more challenging to pinpoint their mechanistic basis. Consequently, less mechanistic interpretation has been performed on AI-based protein structure predictors (70, 71). Nevertheless, a couple very recent studies may provide glimpses of how AI models predict multiple protein conformations.

Late breaking work indicates that both AlphaFold2 and AlphaFold3 predict alternative conformations of some fold switchers by association with sparse sequence patterns (72). These associations occur through their transformer architectures (Evoformer and Pairformer, respectively), which learn structural context by finding relationships between amino acids within a given sequence. The sparse patterns–readily identified from AF2's weights–correspond to positions of one to three amino acids, which are sufficient to switch the conformations of AlphaFold2 and AlphaFold3 models when mutated. These models do not always match experimental ground truth, however, indicating that AlphaFold has not fully learned the features that define alternative conformations of fold switchers. This is consistent with a mechanistic study finding that ESM-2, also based on a transformer architecture, does not require full context to predict ECs (73). Together, these studies show that transformer architectures have limited sensitivities (74), sometimes leading to false associations based on spurious learned correlations. It also may explain why diffusion-based architectures– such as AlphaFold3–perform worse at predicting alternative conformations than expected (30). Though diffusion models were expected to produce more structural diversity by denoising inputs in different ways, this potential may be limited by their upstream transformer models.

Sparse pattern recognition arising from transformer models could also explain why AI-based protein structure predictors fail at tasks beyond alternative protein conformation prediction. For instance, co-folding models, such as AlphaFold3, confidently dock ligands to the same locations within a given structure regardless of its amino acid sequence (36). This unphysical result highlights how these models fail to associate the correct features of a protein (amino acid sequence of the binding pocket) with presence or absence of binding. Further, AlphaFold2 confidently predicts that some repeated intrinsically disordered protein sequences assume well-folded beta-solenoid structures (75). Spurious sequence association may explain this phenomenon also.

## 3. Molecular dynamics (MD) simulations explore conformational landscapes between two known fold-switching conformations

MD simulations offer an orthogonal approach for modeling alternative conformations leveraging physical principles rather than learning from large datasets. Though this principled approach may be more generalizable than a specific training data set, fold-switching proteins are challenging targets because of the slow timescales on which they switch—usually on the order of seconds or slower (8, 63, 76). Thus, to date, molecular dynamics simulations have been used to explore conformational landscapes of fold-switching proteins when both folded conformations are known but have not successfully identified a new alternative fold-switching conformation from first principles. Several simulation techniques are now discussed.

### 3.1 Replica exchange molecular dynamics (REMD)

Replica exchange molecular dynamics (REMD), first introduced by Sugita and Okamoto in 1999, combines MD simulations with Monte Carlo sampling to explore different protein conformations at different temperatures (77). Instead of equilibrium MD simulations, REMD generates the ensembles by combining numerous MD trajectories performed at different temperatures to efficiently explore both low- and high-probability conformations. By periodically exchanging low-probability conformations generated at high temperatures into lower temperature simulations, it is possible to explore the energy landscapes of folded proteins with shorter MD trajectories (78-80). For example, David E. Shaw Research first showed the reversible folding process of WW domains with distinct conformational states through one microsecond of all-atom equilibrium MD simulations (79). Later, they determined the thermodynamics and kinetics of ubiquitin folding pathways through microseconds of all-atom MD simulations (81, 82). Beck et al. explored the different energy landscapes of protein folding via combined REMD and conventional MD simulations (83), while Andrec et al., suggested the folding pathways of the C-terminal peptide from the B1 domain of protein G through REMD and kinetic modeling (84). By contrast, hundred microseconds of equilibrium MD simulations would be required to simulate similar pathways, often a computationally intractable problem.

Recent progress has been made in simulating ensembles of proteins that assume disparate conformations. For instance, an 18 μs REMD simulation with explicit solvent and all-atom conditions showed two distinct minima in the energy landscape corresponding to two different structures of α-synuclein residues 35-97 (85). These simulations indicated synuclein conformations with both compact and extended α-helices, while simulations at higher temperature suggested that more β-hairpin structures were populated. Aggregation mechanisms of amyloid-forming peptides such full-length human Islet polypeptide (hIAPP) and Aβ$_{16-22}$ have also been modeled by observing structure conversion from disordered random coils to β-sheet using REMD simulations (86, 87). Specifically, Choi et al., verified the β-sheet formation mechanism of wildtype and mutated hIAPP dimer through the experimental characterization. Hansmann group reported the structural transition between amyloid oligomer and fibrils (88). They pointed out the different interchain hydrogen bonding from each structure.

Several groups have simulated putative structural ensembles of fold-switching proteins using REMD or variations thereof. The Hansmann group identified and clustered multiple conformations of the C-terminal domain of RfaH and full-length lymphotactin using their replica-exchange with tunneling method (89-92). This method is a hybrid MD/Monte Carlo (MC) with replica exchange, which can efficiently explore the fold switching landscape. Several distinct states–including possible intermediates–of the fold-switching proteins were identified. Further, a recently developed method called modeling employing limited data (MELD) acceleration of MD (MELDxMD) successfully simulated the structural ensembles of the fold-switching GA/GB system and the C-terminal domain of RfaH, which undergoes a dramatic α-helix-to-β-sheet conversion (93); these systems were also simulated using replica exchange with tunneling (94, 95). The conformational ensemble of RfaH's C-terminal domain was also simulated using replica exchange with hybrid tempering (REHT), which controls the temperature of the solvent (water) as well as the solute (96). It should be noted that in all these cases RfaH's CTD was modeled to assume its helical hairpin structure in isolation. NMR experiments indicate that this hairpin is observed in the presence of its N-terminal domain only, however (63, 64, 97).

Replica exchange is a cutting-edge method for exploring conformational changes with all-atom interactions. However, several factors such as the size of the protein and sufficient sampling with explicit solvent conditions require substantial computational power, which can be a barrier for running the REMD. Applying implicit solvent methods or coarse-grained MD simulations (CG-MD) can lower this barrier (98, 99), though that can sacrifice some accuracy. Such limitations make it difficult to sample proteins larger than small domains <100 amino acids. Simulating fold-switching proteins is further confounded by their long exchange timescales on the order of seconds or more (8, 63, 76), which requires more computing power. Further, simulating the conditions that give rise to fold switching can add extra challenges since they can involve binding of large proteins (21) and crowding effects (100), for instance.

### 3.2 Structure-based models

Structural biology has confirmed the existence of fold-switching proteins by resolving their different folded structures, but it does not explain how these proteins switch between folds. To address this, dual-basin structure-based models (SBMs)—simplified computational models grounded in protein folding principles—are used (39). Unlike traditional force fields, dual-basin SBMs simplify and approximate the distribution of energies from native contacts of two different conformations (e.g. dominant and alternative structures of fold-switching proteins). Since SBMs focus on native protein interactions, they may not accurately model a protein's complex structural landscape, which can also involve non-native interactions. Nevertheless, these models have been successfully applied to several proteins, showing that fold-switching typically involves intermediates with native-like structure. In some cases, these intermediates are consistent with experiment (38).

Dual-basin SBMs offer several advantages for studying metamorphic protein refolding. They can be combined with detailed physicochemical force fields to use changes in native contacts as reaction coordinates for enhanced sampling. This approach has revealed mechanisms like hydrogen bond networks driving fold-switching in intermediate states in RfaH's fold-switch (101). These models also effectively track structural changes in complex systems like influenza hemagglutinin (102), SARS-CoV-2 spike proteins (103), and fold-switching tied to quaternary structure changes (e.g., KaiB) or interactions with large complexes (104). A more detailed discussion can be found here (105).

### 3.3 Other MD-based approaches.

Ensembles of several fold-switching proteins have also been simulated with other MD-based approaches. Most notably, the α-helix-to-β-hairpin transition of the plant pathogen PopP2 was simulated with metadynamics simulations (106). These simulations suggested an energy barrier of ~3 kcal/mol between its two experimentally characterized conformations. Further, the open-to-closed conformational transition of the fold-switching tuberculosis protein PimA was simulated using steered MD (107). Increasingly, machine-learning methods are being used to accelerate MD-like simulations (108-111). Though the timescale of these is often too slow to simulate conformational transitions of fold switchers, one of them–Upside–was recently used to simulate some of the fold-switching transition of KaiB (8). As deep learning models continue to improve, we are optimistic more robust modeling of fold-switching trajectories may occur in the not-too-distant future.

## 4. Future opportunities

Though deep learning models have revolutionized protein structure prediction, they continue to struggle to predict alternative protein conformations and protein ensembles by extension. This is especially true for conformations that lack training-set homologs and/or differ substantially from dominant predicted structures (46). Further, assessing when an alternative conformation is predicted accurately is not always straightforward. Sometimes, confident predictions of alternative conformations are inconsistent with experiments (30). Other times, low-confidence predictions of alternative conformations are correct (24). These shortcomings likely arise because deep learning models have not learned protein folding physics (24, 37, 112). Without a biophysical basis, predictions rely on other factors, such as training set prevalence (32) and memorization of uncommon conformations (24, 31). Consequently, the generalizability of deep learning models is limited for fold switchers (30), other sorts of conformational changes (32, 55), protein-peptide interactions (34), and protein-ligand interactions (35, 36).

In closing, we mention some promising areas of advancement. First, deep learning methods have recently been combined with MD simulations to model protein ensembles more accurately (113, 114). The advantage of these methods is that they may not require experimentally determined structures of all populated conformations, especially if those conformations are populated in the training set. Second, high-throughput methods have recently been used to measure protein stabilities (115) and the energy landscapes of some protein families (116). Such approaches may supply AI models with the data required to make more robust predictions, at least in some cases. Third, the importance of temperature in the conformational equilibria of fold-switching proteins is increasingly appreciated (117). Particularly, it has been suggested that some alternative conformations of fold switchers respond to cold denaturation (118). AI-based models informed by physical information, such as energy landscape and temperature, may therefore lead to improved models of alternative conformations and protein ensembles in the future. For instance, Physics-Informed Neural Networks (PINNs) have been applied to protein-small molecule binding (119), rigid-body protein docking (120), and protein folding dynamics (121). Fourthly, though coevolutionary information does not appear to drive AI-based predictions of alternative fold-switched conformations, such information exists in MSAs some of the time (122). Future modeling strategies may leverage this observation to make better predictors. Finally, while current predictors have yielded correct models of homologs of known fold switchers, they are not yet robust enough to predict new fold switchers from genomic sequences. If this challenge is addressed, it may indicate progress in modeling alternative conformations that extends to predicting protein ensembles more generally.

**Acknowledgments**

We thank Devlina Chakravarty for critically reading this manuscript. This work utilized resources from the NIH HPC Biowulf cluster (http://hpc.nih.gov) and was supported by the Division of Intramural Research at the National Library of Medicine, National Institutes of Health (NIH, LM202011, L.L.P.). The contributions of the NIH authors are considered Works of the United States Government. The findings and conclusions presented in this paper are those of the author(s) and do not necessarily reflect the views of the NIH or the U.S. Department of Health and Human Services. This review will be published in the 2026 edition of Annual Reviews of Biomedical Data Science: https://doi.org/10.1146/annurev-biodatasci-092524-114822.